\documentclass[12pt]{article}
\usepackage[reqno]{amsmath}

\usepackage{epsfig}
\usepackage{array}
\usepackage{float}

\usepackage{a4}

\usepackage{a4wide}
\usepackage{wasysym}

\def\Jo#1#2#3#4{#1 {\bf #2}, #3 (#4)}

\def\NPB{Nucl.\ Phys.\ B}

\def\PLB{Phys.\ Lett.\ B}
\def\PRL{Phys.\ Rev.\ Lett.\ }
\def\PRD{Phys.\ Rev.\ D}

\def\ZPC{Z.\ Phys.\ C}
\def\IJMP{Int.\ J.\ of Mod.\ Phys.\ A}
\def\JHEP{JHEP}

\def\ra{\rightarrow}
\def\be{\begin{equation}}
\def\ee{\end{equation}}
\def\gs{\mathrel{
   \rlap{\raise 0.511ex \hbox{$>$}}{\lower 0.511ex \hbox{$\sim$}}}}
\def\ls{\mathrel{
   \rlap{\raise 0.511ex \hbox{$<$}}{\lower 0.511ex \hbox{$\sim$}}}}

\newcommand{\ba}{\begin{array}{c}}
\newcommand{\baz}{\begin{array}{cc}}
\newcommand{\bad}{\begin{array}{ccc}}
\newcommand{\bea}{\begin{equation} \begin{array}{c}}
\newcommand{\eea}{ \end{array} \end{equation}}
\newcommand{\ea}{\end{array}}
\newcommand{\D}{\displaystyle}
\newcommand{\dms}{\mbox{$\Delta m^2_{\odot}$}}
\newcommand{\dma}{\mbox{$\Delta m^2_{\rm A}$}}
\newcommand{\meff}{\mbox{$\langle m \rangle$}}

\begin{document}
\title{
\hfill {\small Ref.\ SISSA 96/2003/EP}\\
\hfill {\small hep-ph/0311142} \\ \vskip 1cm
\bf
Hierarchical Matrices in the See-Saw Mechanism, large Neutrino Mixing and  
Leptogenesis}
\author{
Werner Rodejohann\footnote{Email: {\tt werner@sissa.it}}\\[0.3cm] 
{\normalsize \it Scuola Internazionale Superiore di Studi Avanzati,
I-34014 Trieste, Italy}\\
{\normalsize and }\\
{\normalsize \it Istituto Nazionale di Fisica Nucleare,
Sezione di Trieste, I-34014 Trieste, Italy}\\}
\date{}
\maketitle
\thispagestyle{empty}
\begin{abstract}
\noindent We consider the see--saw mechanism for hierarchical 
Dirac and Majorana neutrino mass matrices $m_D$ and $M_R$, 
including the $CP$ violating phases. 
Simple arguments about the structure of the neutrino mass 
matrix and the requirement of successful leptogenesis lead to the 
situation that one of the right--handed Majorana neutrinos 
is much heavier than the other two, which in turn display a 
rather mild hierarchy. 
It is investigated how for the neutrino mixing one small and two 
large angles are generated. The mixing matrix element 
$|U_{e3}|^2$ is larger than $10^{-3}$ 
and a characteristic 
ratio between the branching ratios of lepton flavor violating 
charged lepton decays $\ell_j \ra \ell_i  \, \gamma$ is found. 
Successful leptogenesis implies sizable $CP$ violation 
in oscillation experiments. 
As in the original minimal see--saw model, 
the signs of the baryon asymmetry of the universe and of 
the $CP$ asymmetry in neutrino oscillations are equal and there is no 
connection between the leptogenesis phase and the effective mass as 
measurable in neutrinoless double beta decay.

\end{abstract}

\newpage

\section{\label{sec:intro}Introduction}
The fact that two mixing angles in the neutrino mixing matrix are 
large \cite{oscrev} is a severe difference with respect to the quark sector. 
In the latter, hierarchical mass matrices are the most natural 
explanation for small mixing angles. Thus, it is natural to assume that 
in a GUT framework also the 
Dirac mass matrix $m_D$ and the Majorana mass matrix $M_R$, both appearing 
in the see--saw mechanism \cite{seesaw}, are of 
hierarchical structure, i.e., of close to diagonal form. 
In the see--saw mechanism the neutrino mass 
matrix $m_\nu$ is a matrix product containing $m_D$ and $M_R$. 
Consequently, it 
is possible that $m_\nu$ does not display 
a close to diagonal structure\footnote{The names 
``see--saw enhancement'' or ``correlated hierarchy'' are sometimes 
given to this phenomenon.}, in contrast to the fundamental matrices 
$m_D$ and $M_R$ \cite{earlier}.  
Accordingly, the observed neutrino mixing can take the 
characteristic form with two large angles and one small one. 
The purpose of the present note is to readdress this point 
including effects of the $CP$ phases and investigate its consequences for 
leptogenesis and for the branching ratios of 
lepton flavor violating (LFV) charged lepton decays like 
$\mu \ra e \, \gamma$. 
In order 
to reach a hierarchical mass spectrum, the 23 block of $m_\nu$ has to 
be approximately degenerate with entries larger than 
the remaining elements \cite{dominant,king,ich}. 
Working within useful parameterizations of $m_D$ and $M_R$, 
these requirements lead to the possibility that 
one of the right--handed Majorana neutrinos is much heavier than the 
other two. Successful leptogenesis then implies a rather mild 
hierarchy between the latter. 
In this simple framework one can obtain neutrino mixing 
phenomenology in accordance with data, predicts $|U_{e3}|^2 \gs 10^{-3}$ 
and finds a characteristic ratio of the branching ratios of the 
LFV charged lepton 
decays. The baryon asymmetry of the universe and the $CP$ asymmetry 
measurable in neutrino oscillations are directly connected, since they 
depend in the same way on the same phase. 
No connection between the leptogenesis phase and 
the effective mass as testable in neutrinoless double beta decay 
is present. 
The model under study is in this sense very similar to the  
minimal see--saw model \cite{FGY1}, which contains only two heavy 
Majorana neutrino and two zeros in the Dirac mass matrix.\\ 

\noindent In Section \ref{sec:frame} we will shortly review the 
formalism of neutrino mixing and leptogenesis. We investigate how 
hierarchical Dirac and Majorana mass matrices lead to large 
neutrino mixing in a simplified $2 \times 2$ case in Section \ref{sec:22}. 
The realistic $3 \times 3$ case is treated in Section \ref{sec:33}, 
where also the predictions for 
leptogenesis and low energy observables are investigated. We 
conclude in Section \ref{sec:concl}.

\section{\label{sec:frame}Framework}
The neutrino mass matrix is given by the see--saw mechanism 
\cite{seesaw} as 
\be 
\label{eq:seesaw}
m_\nu \simeq - m_D \, M_R^{-1} \, m_D^T~, 
\ee
where $m_D$ is a Dirac mass matrix and $M_R$ the mass matrix of the 
right--handed Majorana neutrinos. We shall work in a basis in which 
both the charged lepton mass matrix and $M_R$ are real and diagonal, i.e., 
$M_R = {\rm diag}(M_1, M_2, M_3)$ with real $M_3 > M_2 > M_1$. 
The largest mass $M_3$ is expected to lie around or below the 
unification scale $M_{\rm GUT} \simeq 10^{16}$ GeV\@.  
The matrix $m_\nu$ is observable in terms of 
\be \label{eq:mnu}
m_\nu = U^\dagger \, m_\nu^{\rm diag} \, U^\ast~.
\ee
Here $m_\nu^{\rm diag}$ is a diagonal matrix containing the 
light neutrino mass eigenstates $m_i$ and $U$ is the unitary 
Pontecorvo--Maki--Nagakawa--Sakata 
\cite{PMNS} lepton 
mixing matrix, which can be parametrized as 
\be \label{eq:UPMNS}
U = O_{23} \, O_{13}^\delta \, O_{12} \, P~. 
\ee
$O_{ij}$ are rotation matrices, e.g., 
\be
O_{13}^\delta = \left( \bad 
c_{13} & 0 & s_{13} \, e^{i \delta} \\[0.3cm]
0 & 1 & 0  \\[0.3cm]
-s_{13} \, e^{-i \delta} & 0 & c_{13} \ea \right)~, 
\ee
where $c_{13} = \cos \theta_{13}$, $s_{13} = \sin \theta_{13}$ and 
$\delta$ is the ``Dirac phase'' measurable in neutrino oscillations. 
The matrices $O_{12}$ and $O_{23}$ are 
real and $P$ is a diagonal phase matrix containing the 
two additional Majorana phases. In total, 
\bea \label{eq:Upara}
U = \left( \bad 
c_{12} c_{13} & s_{12} c_{13} & s_{13}  \\[0.2cm] 
-s_{12} c_{23} - c_{12} s_{23} s_{13} e^{i \delta} 
& c_{12} c_{23} - s_{12} s_{23} s_{13} e^{i \delta} 
& s_{23} c_{13} e^{i \delta}\\[0.2cm] 
s_{12} s_{23} - c_{12} c_{23} s_{13} e^{i \delta} & 
- c_{12} s_{23} - s_{12} c_{23} s_{13} e^{i \delta} 
& c_{23} c_{13}e^{i \delta}\\ 
               \ea   \right) 
 {\rm diag}(1, e^{i \alpha}, e^{i \beta }) \, . 
\eea
Observation from previous experiments \cite{oscrev} 
as well as inclusion of the recent SNO salt phase data \cite{SNO} 
implies the following values of 
the oscillation parameters \cite{valle}, given at $3 \sigma$: 
\bea 
\tan 2 \theta_{12} \simeq 1.5 \ldots 4.4 ~,\\[0.3cm]
\tan 2 \theta_{13} \ls  0.45 ~,\\[0.3cm]
|\tan 2 \theta_{23}| \gs  2~,\\[0.3cm]
\dms \simeq (5.4 \ldots 9.5) \cdot 10^{-5} \, 
\rm eV^2~,\\[0.3cm] 
\dma \simeq (1.4 \ldots 3.7) \cdot 10^{-3} \, \rm eV^2~. 
\eea
Typical best--fit points are $\tan^2 \theta_{12} = 0.45$ and 
$\theta_{23} = \pi/4$, corresponding to $\tan 2 \theta_{12} \simeq 2.4$ 
and $\tan 2 \theta_{23} \gg 1$. We have therefore two large and 
one small mixing angle, in sharp contrast to the situation 
present in quark mixing.\\

\noindent The presence of heavy right--handed Majorana 
neutrinos in the see--saw mechanism means that the possibility 
of leptogenesis \cite{leptogenesis} is included. Thus, the see--saw mechanism 
gains a large amount of attractiveness. 
Leptogenesis explains the 
baryon asymmetry of the universe through the $CP$ asymmetric 
out--of--equilibrium decay of heavy right--handed 
Majorana neutrinos occurring much before 
the electroweak phase transitions. It is governed by the decay 
asymmetry \cite{leptogenesis,leptorev}
\be \label{eq:eps}
\ba 
\varepsilon_1 
\simeq  \D \frac{\D 1}{\D 8 \, \pi \, v^2} \frac{\D 1}{(m_D^\dagger m_D)_{11}} 
\sum_{j \neq i} {\rm Im} (m_D^\dagger m_D)^2_{1j} \, f(M_j^2/M_1^2)~,
\ea 
\ee
where $f(x)$ is a function whose limit for $x \gg 1$, i.e., 
hierarchical neutrinos\footnote{We shall not 
discuss the possibility of degenerate Majorana neutrinos, 
whose decay asymmetry is resonantly enhanced \cite{res}.}, 
is $-3/\sqrt{x}$. Values of  
$|\varepsilon_1| \gs 10^{-7}$ and 
$M_1 \gs 10^{9}$ GeV are required in order to produce a sufficient 
baryon asymmetry \cite{leptorev,washout}. 
There is a tendency of this lower mass limit 
to be in conflict with bounds on the 
reheating temperature, which stem from the requirement that the decay 
products of the gravitino do not spoil Big Bang Nucleosynthesis 
predictions. From this condition one finds upper limits of less than 
$M_1 \ls 10^9 \ldots 10^{10}$ GeV \cite{gravitino}.

\noindent The baryon asymmetry is 
positive when $\varepsilon_1$ is negative, 
because it holds $Y_B \propto c \, \varepsilon_1$ \cite{leptorev}, 
where $c$ is a negative constant stemming from the conversion of 
the lepton asymmetry into a baryon asymmetry.

\section{\label{sec:22}$2 \times 2$ Case}
We shall analyze the generation of large mixing in $m_\nu$ from 
hierarchical $m_D$ and $M_R$ first in a simplified $2 \times 2$ framework.  
Consider a complex symmetric matrix 
\be
m = 
\left( 
\baz 
a & b \\[0.3cm]
b & d
\ea 
\right)~, 
\ee
which is diagonalized by a unitary matrix $U$ through  
\bea
m^{\rm diag} = 
\left( 
\baz 
m_1 & 0 \\[0.3cm]
0 & m_2 
\ea 
\right) = 
U^T \, m  \, U~, \, \mbox{ where } \,  
U = 
\left( 
\baz 
\cos \theta & \sin \theta \, e^{i \phi} \\[0.3cm]
- \sin \theta \, e^{-i \phi} & \cos \theta
\ea 
\right)~.
\eea
In general, a symmetric matrix $2 \times 2$ is diagonalized by $U P$, 
where $U$ is given above and $P$ is a diagonal phase matrix. 
By redefining the charged lepton fields, these two additional phases 
can be absorbed. 
The eigenvalues $m_1$ and $m_2$ with $m_2 > m_1$ are trivial to obtain.  
The mixing angle $\theta$ is given by the equation 
\be
\tan 2 \theta = \frac{2 \, b }{d \, e^{-i \phi} - a \, e^{i \phi}}~.  
\ee
The phase $\phi$ is defined by the 
requirement of the angle $\theta$ being real, i.e., 
\be
\arg (b) = \arg(d \, e^{-i \phi} - a \, e^{i \phi})~.
\ee

\noindent Now consider in a simple $2 \times 2$ case 
hierarchical Dirac and Majorana mass matrices, i.e., 
\be
m_D = m  
\left( 
\baz 
\epsilon_D^2 & A \, \epsilon_D \\[0.3cm]
B \, \epsilon_D & 1 
\ea
\right) \mbox{ and } 
M_R = M  
\left( 
\baz 
\epsilon_M & 0 \\[0.3cm]
0 & 1 
\ea
\right)~, 
\ee 
with $\epsilon_D, \epsilon_M \ll 1$ but an unspecified hierarchy between 
$\epsilon_D$ and $\epsilon_M$. 
The complex coefficients $A = a \, e^{i \alpha}$ and $B = b \, e^{i \beta}$ 
with real $a$ and $b$ have absolute values of order one.  
Inserting the matrices in the see--saw formula (\ref{eq:seesaw}) yields 
\be \label{eq:mnu22}
m_\nu = - \frac{m^2}{M}  
\left( 
\baz 
\frac{\D \epsilon_D^4}{\D \epsilon_M} + A^2 \, \epsilon_D^2 & 
A \, \epsilon_D + B \, \frac{\D \epsilon_D^3}{\D \epsilon_M} \\[0.3cm]
 \cdot & 
1 + B^2 \, \frac{\D \epsilon_D^2}{\D \epsilon_M}
\ea 
\right) = 
- \frac{m^2}{M}  
\left( 
\baz 
\epsilon_D^2 \, ( A^2 + \eta) & 
\epsilon_D \, (A + B \, \eta) \\[0.3cm]
 \cdot & 
1 + B^2 \, \eta^2
\ea 
\right)
~,
\ee
where we defined the characteristic quantity 
$\eta \equiv \epsilon_D^2/\epsilon_M$. The magnitude of the 
mixing angle is therefore governed by the ratio of the hierarchies of the 
Dirac and Majorana masses. Namely: 
\be \label{eq:t2t22}
\tan 2 \theta = 2 \, \epsilon_D \,
\frac{ A + \eta \, B }
{1 + \eta \, (B^2 - e^{2i \phi} \, A^2 \, \epsilon_M - 
e^{2 i \phi} \, \epsilon_D^2)}~e^{i \phi}
~.
\ee
From Eq.\ (\ref{eq:t2t22}) one encounters several interesting 
special cases, some of which are discussed in the following:\\ 
\noindent$1)$ $\eta \simeq 1$ but $\epsilon_{M,D} \ll 1$: similar 
hierarchy in $m_D$ and $M_R$

\noindent Then, we find for the mass matrix and the mixing angle 
\be \label{eq:tmax22}
\bad  
m_\nu \simeq -\frac{\D m^2}{\D M} 
\left( 
\baz 
0 & \epsilon_D \, (A + B) \\[0.3cm]
\cdot & 1 + B^2 
\ea 
\right) & \Rightarrow &   
\tan 2 \theta \simeq 2 \, \epsilon_D \, 
\sqrt{\frac{\D a^2 + b^2 + 2 \, a \, b \, c_{\alpha - \beta}}
{\D 1 + b^4 + 2 \, b^2 \, c_{2\beta}}}~.
\ea 
\ee
Values of $\beta \simeq \pi/2$ and $b \simeq 1$ can thus lead to 
(close--to--)maximal mixing as observed in the atmospheric neutrino 
oscillation experiments. In this case, $\phi \simeq -\arg(A + i)$. 
Also, relaxing the conditions for $b$ and $\beta$ a bit can 
lead to the observed large but not maximal mixing in solar neutrino 
oscillation experiments.\\  

\noindent$2)$ $\eta \ll 1$: stronger hierarchy in $m_D$

\noindent The mass matrix and mixing are now given by
\be
\bad  
m_\nu \simeq -\frac{\D m^2}{\D M} 
\left( 
\baz 
0 & A \, \epsilon_D \\[0.3cm]
\cdot & 1 
\ea 
\right) & \Rightarrow &
\tan 2 \theta \simeq 2 \, \epsilon_D \, a~,
\ea
\ee
which, for large but still reasonable choices of  
$\epsilon_D \simeq \sin \theta_C \simeq 0.22$ and $a \gs 4$ 
yields $\tan 2 \theta \gs \sqrt{3}$, i.e., $\theta \gs \pi/6$, 
as implied by the observed non--maximal large 
mixing in the solar neutrino oscillation experiments. 
More naturally, smaller values 
of $\epsilon_D$ and $a$ can easily reproduce the small 
mixing parameter as implied 
by the CHOOZ and Palo Verde reactor neutrino oscillation experiments. 
For the phase holds $\phi \simeq -\alpha$.\\

\noindent $3)$ $\eta \gg 1$: stronger hierarchy in $M_R$

\noindent The mixing is found to be
\be
\bad  
m_\nu \simeq -\frac{\D m^2}{\D M} 
\left( 
\baz 
0 & B \, \epsilon_D \eta \\[0.3cm]
\cdot & B^2 \, \eta 
\ea 
\right) & \Rightarrow &
\tan 2 \theta \simeq 2 \, \epsilon_D \, \frac{\D 1}{\D b}~,
\ea
\ee
for which similar arguments as for the case $\eta \ll 1$ hold. The 
phase is given by $\phi \simeq \beta$.\\ 

\noindent To sum up, hierarchical Dirac and Majorana 
mass matrices reproduce for specific choices of the hierarchies 
and parameters all observed types of neutrino mixing, 
(close--to--)maximal, non--maximal large and small mixing. 
Exactly maximal and vanishing mixing requires some fine--tuning. 
Vanishing mixing would be obtained for  
$|A + \eta \, B| \simeq 0$ or equivalently 
$a^2 + b^2 \, \eta^2 = - 2 \, a \, b \, c_{\alpha - \beta}$. 
We show in Fig.\ \ref{fig:examix22} several examples of the 
mixing obtained with specific choices of $\epsilon_D$, $A$ 
and $B$. One finds from the figure and the discussion in this Section that 
in order to obtain (close--to--)maximal mixing there is --- in the given 
parametrization --- a crucial dependence on the hierarchies of the 
fundamental matrices $m_D$ and $M_R$. Also the phases 
play an important role. Leptogenesis in turn requires  
the presence of $CP$ 
violation\footnote{Note though that {\it in general} no link between 
low and high energy 
$CP$ violation exists \cite{branco,PPR3} and any such connection 
will be model dependent.} and --- from Eq.\ (\ref{eq:eps}) --- 
depends on $m_D$ and $M_R$, therefore also on the 
ratio of the hierarchies. 
We should thus analyze leptogenesis in this scenario. 
The decay asymmetry reads 
\be
\varepsilon_1 = \frac{3 \, \epsilon_M }{4 \, \pi} \, \frac{m^2}{v^2} \, 
\frac{1}{b^2 + \epsilon_D^2} 
\left( (a \, \epsilon_D^2 \, \cos \alpha + b \, \cos \beta) \, 
(b \, \sin \beta - a \, \epsilon_D^2 \, \sin \alpha ) 
\right) \simeq 
\frac{3\, \epsilon_M }{8 \, \pi} \, \sin 2 \beta~,
\ee
where terms of order $\epsilon_D^2$ were neglected and 
$m \simeq v$ was used. We can construct a very interesting special case: 
suppose that the mass matrix parameters take the values 
$b \simeq 1$, $\epsilon_D \simeq 0.1$ and $\eta \simeq 1$. 
Then, from Eq.\ (\ref{eq:tmax22}), we see that 
maximal mixing is only possible for $\beta \simeq \pi/2$. For 
this value of the phase, however, the decay asymmetry is highly suppressed. 
Therefore, maximal mixing implies a too small baryon asymmetry, or in 
other words, requiring a non--zero baryon asymmetry implies 
non--maximal neutrino mixing.  We shall encounter a slightly 
similar effect in the next Section for the $3 \times 3$ case. 
Stressed is here that the same $CP$ phase can affect the magnitude of 
neutrino mixing angles and the value of the 
baryon asymmetry of the universe.

\section{\label{sec:33}The $3 \times 3$ case}
Let us turn now to the appropriate 3 flavor case.  
We can parametrize the relevant mass matrices $m_D$ and $M_R$ now as 
\be \label{eq:md}
m_D \simeq m 
\left( 
\bad 
0 & A \, \epsilon_D^3 & 0 \\[0.3cm] 
B \, \epsilon_D^3 & \epsilon_D^2 & F \, \epsilon_D^2 \\[0.3cm] 
0 & g \, \epsilon_D^2 & 1 
\ea
\right)\, , \, 
M_R = M 
\left( 
\bad 
\epsilon_{M1} & 0 & 0 \\[0.3cm] 
0 & \epsilon_{M2} & 0 \\[0.3cm] 
0 & 0 & 1
\ea
\right)~.
\ee
For later use we define $A = a \, e^{i \alpha}$, 
$B = b \, e^{i \beta}$ and $F = f \, e^{i \phi}$; $g$ can be chosen real. 
Again, the complex coefficients have absolute values 
of order one, so has $g$.   
Small entries in the 11, 13 and 31 elements of $m_D$ are neglected 
(see below) and it holds 
$\epsilon_{M1} < \epsilon_{M2}$. We choose now the following 
parameters describing the relative hierarchy in $m_D$ and $M_R$: 
\be
\eta_1 = \epsilon_D^4/\epsilon_{M1} ~\mbox{ and }~ 
\eta_2 = \epsilon_D^4/\epsilon_{M2} ~\mbox{ with }~ \eta_1 > \eta_2~. 
\ee
Let us choose a typical expansion parameter in $m_D$ of 
$\epsilon_D \simeq 0.1$ and an overall mass scale 
$m \simeq v \simeq 174$ GeV\@. 
Using the see--saw formula we find for $m_\nu$:
\be \label{eq:mnu33} 
m_\nu \simeq \frac{-m^2}{M} 
\left( 
\bad 
A^2 \, \epsilon_D^2 \, \eta_2 & A \, \epsilon_D \, \eta_2 & 
A \, g \, \epsilon_D \, \eta_2 \\[0.3cm]
\cdot & \eta_2 + B^2 \epsilon_D^2 \, \eta_1 
+ F^2 \epsilon_D^4 & 
F \, \epsilon_D^2 + g \, \eta_2 \\[0.3cm]
\cdot & \cdot & 1 + g^2 \, \eta_2 
\ea
\right)~.
\ee
\noindent The light neutrino mass scheme will of course be hierarchical. 
To have an approximately  
degenerate spectrum in the 23 submatrix of $m_\nu$ 
(with scale $\sim \sqrt{\dma}$) it is required that $g \simeq 1$ and 
$\eta_2 \simeq 1$ or $\eta_2 \simeq 10$. Larger values are incompatible 
with $m \simeq v$ and $M \ls 10^{16}$ GeV\@. 
Later on it will be shown that $\tan 2 \theta_{12}$, 
where $\theta_{12}$ is the mixing angle    
governing the solar neutrino oscillations, is proportional to 
$\epsilon_D \, \eta_2$ and thus the larger value of $\eta_2 \simeq 10$ 
is implied. 
Thus, $\epsilon_{M2} = \epsilon_D^4/\eta_2 
\simeq 10^{-5}$, i.e., the heaviest 
Majorana neutrino has a much larger mass than the other two. 

\noindent We can gain even more insight in the hierarchy of 
$M_R$ by looking at the 
decay asymmetry of the heavy Majorana neutrinos. It reads 
\bea \label{eq:vareps3}
\varepsilon_1 = \frac{\D 3 \, m^2}{\D 8 \, \pi \, v^2} \, \epsilon_D^4 \, 
\left(\frac{\D \epsilon_{M1}}{\D \epsilon_{M2}} \sin 2 \beta  
+ f^2 \, \epsilon_{M1} \, \sin 2(\beta - \phi) \right) \\[0.3cm]
\simeq 0.1 \, \epsilon_D^4 \, 
\left(\frac{\D \epsilon_{M1}}{\D \epsilon_{M2}} \sin 2 \beta  
+ f^2 \, \epsilon_{M1} \, \sin 2(\beta - \phi) \right)\\[0.3cm]
\simeq 0.1 \, \epsilon_D^4 \, 
\frac{\D \epsilon_{M1}}{\D \epsilon_{M2}} \sin 2 \beta~, 
\eea
where we used $\epsilon_{M1} \ll 1$ and assumed again $m \simeq v$. 
We can identify the leptogenesis phase $\beta$. 
Since the decay asymmetry should be negative, 
we can constrain $\beta$ to lie between $\pi/2$ and $\pi$ or 
between $3\pi/2$ and $2\pi$. 
In order to reach a favorable 
value of $|\varepsilon_1| \gs 10^{-7}$, the factor 
$\epsilon_{M2}/\epsilon_{M1} = \eta_1 /\eta_2$ should not exceed 
$\sim 10$. Therefore, the two lightest Majorana neutrinos display 
a rather mild hierarchy.  
The requirements for the 
structure of $m_\nu$ and successful leptogenesis 
therefore determine the hierarchy of $M_R$. 

\noindent For numerical estimates of the 
obtained quantities we shall use in the following the representative values  
$\epsilon_{M1} = 10^{-6}$, $\epsilon_{M2} = 10^{-5}$ 
and $\epsilon_{D} = 0.1$. 
These choices basically eliminate the parameter 
$F = f \, e^{i \phi}$ from the problem. The ratios of the branching 
ratios of the LFV violating charged lepton decays in Eq.\ (\ref{eq:BRs}) 
remain however somewhat sensitive to this parameter. 
Looking with the given parameter set for $\epsilon_D$, $\epsilon_{M1}$ 
and $\epsilon_{M2}$ at Eq.\ (\ref{eq:mnu33}), one notes that 
the terms including $A$ and thus $\alpha$ are subleading. 
One can therefore expect the phase $\beta$ to play the major role in the  
observables under study. We shall see that this is indeed the case.\\

\noindent For thermal leptogenesis the important 
effective mass parameter is given by 
\be
\tilde{m}_1 = \frac{(m_D^\dagger m_D)_{11}}{M_1} \simeq 
\frac{m^2}{M} \, b^2 \, \eta_1 \, \epsilon_D^2~,
\ee
being of the order of the entries in $m_\nu$ and thereby guaranteeing 
for the baryon asymmetry a not too  
strong wash--out factor $\kappa$ (stemming from lepton number 
violating scattering processes) of $\kappa \sim 0.1 - 10^{-3}$ \cite{washout}. 

\noindent We can get a lower limit on the heavy neutrino masses by 
comparing  our 
formula for $\varepsilon_1$ with its analytical upper limit, 
which reads \cite{ibarra}
\be
|\varepsilon_1 | \ls \frac{3}{8 \, \pi \, v^2} \, M_1 \, \sqrt{\dma}~.
\ee
With $\dma \gs 10^{-3}$ eV$^2$ one finds 
\be
M_1 \gs  \epsilon_D^4 \, 
\frac{\D \epsilon_{M1}}{\D \epsilon_{M2}} \, 10^{15} \,~\rm GeV~.
\ee
Therefore,  
for our chosen parameters of 
$\epsilon_D \simeq 0.1$ and $\epsilon_{M1}/\epsilon_{M2} \simeq 0.1$,  
it holds $M_1 \gs 10^{10}$ GeV.\\

\noindent We can now take a closer look at the rates 
of the LFV violating charged lepton decays. 
Assumption of universality of the slepton mass matrices at the GUT 
scale leads via radiative corrections to non--diagonal entries at low 
scale, which give rise to LFV violating charged lepton decays such as 
$\mu \rightarrow e + \gamma$,
$\tau \rightarrow \mu + \gamma$ and 
$\tau \rightarrow e + \gamma$ \cite{lfv}. 
The branching ratios for the decay $\ell_j \ra \ell_i \, \gamma$ 
with $\ell_{(3,2,1)} = \tau, \mu , e$ are approximately 
proportional to $|(m_D m_D^\dagger)_{ji}|^2$. 
In our case, their magnitude is governed by  
\be
BR(\mu \ra e \, \gamma) \propto \left|(m_D m_D^\dagger)_{21}\right|^2
\simeq a^4 \, m^4 \, \epsilon_D^{10}  
\ee
and their ratios are predicted to be 
\be \label{eq:BRs}
BR(\mu \ra e \, \gamma) 
\simeq \frac{1}{g^2} \, BR(\tau \ra e \, \gamma) 
\simeq \frac{a^2}{f^2} \, \epsilon_D^{6} \, 
BR(\tau \ra \mu \, \gamma)~.
\ee
This relation gets modified by the presence of small entries in 
$m_D$, see Section \ref{sec:small}. 

\subsection{\label{sec:diag33}Diagonalization}
As seen, our simple arguments lead 
to the situation in which one of the right--handed 
Majorana mass is much heavier than the other two, which in turn 
display a mild hierarchy. In order to compare our framework with 
the neutrino data, we shall next 
diagonalize the resulting mass matrix $m_\nu$, leaving the definitions and 
details to the Appendix. 
We did not consider the renormalization of the mass matrix since the 
corrections to neutrino masses and mixings are subleading  
in the case of a hierarchical mass spectrum \cite{radcor}, which  
we are considering.\\  

\noindent Observation requires large mixing in the 
23 sector of the matrix $m_\nu$ in Eq.\ (\ref{eq:mnu33}), which is 
given by  
\be \label{eq:m23}
m_\nu^{23} \simeq  \frac{-m^2}{M}
\left(
\baz \eta_2 + B^2 \, \epsilon_D^2 \, \eta_1 & g \, \eta_2 \\[0.3cm]
\cdot & 1 + g^2 \, \eta_2 
\ea
\right) \simeq \frac{-m^2}{M} \, \eta_2 
\left(
\baz 1   & g  \\[0.3cm]
\cdot & g^2  
\ea
\right)
\ee
and diagonalized by the mixing angle 
\be 
\tan 2 \theta_{23} \simeq 
\frac{2 \, g }{g^2 - 1} ~.
\ee
Note that the hierarchy chosen in this analysis renders the 23 submatrix 
quasi real, thereby simplifying the diagonalization procedure, see the 
Appendix for details. 
In order to guarantee a large solar mixing, 
the determinant of $m_\nu^{23}$ should be small \cite{dominant,king}, 
which leads from Eq.\ (\ref{eq:m23}) 
to $|1 + b^2 \, g^2 \, \epsilon_D^2 \, \eta_1 \, e^{2 i \beta}| \ls 1$.

\noindent The deviation from maximal mixing is of order 
\be \label{eq:devatm}
1 - \sin^2 2 \theta_{23} \simeq 
\left( \frac{1 - g^2}{1 + g^2} \right)^2 ~.
\ee 
The largest eigenvalue of $m_\nu^{23}$ is  
\be \label{eq:m3s}
m_3' 
\simeq \frac{-m^2}{\, M} \, \eta_2 \, (1 + g^2) ~. 
\ee
Note that $m_3'$ will not be changed significantly 
by the following two rotations, $m_3' \simeq m_3$, 
and can therefore already be confronted with 
$\sqrt{\dma} \simeq 0.05$ eV\@. Values of $m \simeq v$ 
and $M \simeq 10^{16}$ GeV lead to the desired value if 
$g \simeq 1 $ and $\eta_2 \simeq 10$. 

\noindent It is now straightforward to extend the diagonalization procedure 
from Section \ref{sec:22} in order to obtain the remaining mass 
and mixing parameters. See the Appendix for details. 
One finds for the angle $\theta_{13}$ that 
\be \label{eq:t13}
\tan 2\theta_{13} \simeq \sqrt{2} \, a \, \epsilon_D \, 
\frac{1 + g}{1 + g^2}~, 
\ee 
while the solar neutrino oscillations are triggered by 
\be \label{eq:t12}
\tan 2 \theta_{12} \simeq 
\frac{\sqrt{2} \, a \, \epsilon_D \, \eta_2  \, (1 - g) \, (1 + g^2)} 
{\sqrt{1 + b^2 \, g^2 \, \epsilon_D^2 \, \eta_1  
(b^2 \, g^2 \, \epsilon_D^2 \, \eta_1 + 2 \, c_{2\beta})}}~.
\ee
One notes that $\theta_{13}$ 
is naturally small, $\tan 2 \theta_{13} \propto \epsilon_D$, while 
$\tan 2 \theta_{12}$ is larger than $\tan 2 \theta_{13}$ by approximately 
a factor of $\sim \eta_2$.  
We therefore observe a hierarchy in the mixing angles of the form
\be  \label{eq:resmix}
\tan 2 \theta_{23} \propto \frac{1}{1 - g^2} > \tan 2 \theta_{12} \propto 
\epsilon_D \, \eta_2 > \tan 2 \theta_{13} \propto \epsilon_D~,
\ee 
which is exactly the situation implied by neutrino phenomenology. 
It is seen that, for $\epsilon_D \simeq 0.1$, a value $\eta_2 \sim 10$ is 
required in order to reproduce the large solar neutrino mixing angle, 
which justifies our choice for $\eta_2$ as discussed above. 
Note that the dominator in Eq.\ (\ref{eq:t12}) 
should be smaller than one. In fact, the denominator can be 
identified with 
$|1 + b^2 \, g^2 \, \epsilon_D^2 \, \eta_1 \, e^{2 i \beta}|$, 
and the condition that this quantity is smaller than one was exactly 
the condition to make the determinant of the 23 submatrix of $m_\nu$ 
small. With our assumptions about the hierarchy parameters 
we can make the denominator very small for $b \simeq 1$ 
and $\beta \simeq \pi/2$. This value of $\beta$, however, 
leads via Eq.\ (\ref{eq:vareps3}) to a too small  
baryon asymmetry. We have therefore an interplay 
between the baryon asymmetry of the universe and the non--maximality 
of $\theta_{12}$, which resembles the situation mentioned for the 
$2 \times 2$ case and discussed at the end of Section \ref{sec:22}. 

\noindent Regarding $\theta_{13}$, useful estimates can be performed. 
First of all, one can expect $\theta_{13}$ to be non--zero, because 
$a = 0$ will lead to a too small solar neutrino mixing. 
More precisely, we have for $g \simeq 1$ the estimate 
\be \label{eq:ue3} 
|U_{e3}|^2 \simeq \frac{a^2 \, \epsilon_D^2}{2} \sim (10^{-3} - 10^{-2})~,  
\ee
where we assumed $a$ between 0.5 and 3 and $\epsilon_D = 0.1$. 
These values can be tested in the not too far future \cite{t13}. The 
magnitude of $U_{e3}$ is a crucial prediction for 
neutrino mass models, see, e.g., \cite{Ue3}.

\noindent Fig.\ \ref{fig:examix33} shows for $\epsilon_D = 0.1$, 
$\epsilon_{M1} = 10^{-6}$ and $\epsilon_{M2} = 10^{-5}$ 
the mixing parameter $\tan^2 \theta_{12}$ as obtained from 
Eq.\ (\ref{eq:t12}) for specific choices of $a$, $b$ and 
$g$ as a function of the leptogenesis phase $\beta$. 
The values of $\theta_{23}$ are close to maximal and of 
$\sin^2 \theta_{13}$ close to $10^{-2}$ for all cases plotted,  
confirming our quantitative statements from above. 
Also shown is --- when negative --- the decay asymmetry 
$\varepsilon_1$ from Eq.\ (\ref{eq:vareps3}) 
multiplied with $-10^5$. Its value is of the 
required magnitude for the solar neutrino mixing angle inside its 
experimental range, the angle $\theta_{13}$ below its upper limit and 
atmospheric mixing sufficiently large. Note that too large 
$\tan^2 \theta_{12}$ can lead to a too small decay asymmetry.

\noindent The two remaining mass eigenvalues are complicated functions of 
the parameters $\eta_1$, $\eta_2$, $\epsilon_D$, 
$a$, $b$, $g$, $\alpha$ and $\beta$. We saw above 
that for $\eta_2 \simeq 10$ and 
$M \simeq 10^{16}$ GeV the favorable value of 
$m_3 \simeq \sqrt{\dma}$ is achieved. With this choice for 
$M$, the common factor of $m_{1,2}$ 
is $m^2/M \simeq 3 \cdot 10^{-3}$ eV, which, when multiplied 
with a sum and difference of two 
terms of order one, can, admittedly involving some tuning, 
result in the required 
values of $|m_2|^2 - |m_1|^2 = \dms$. 
For later use we define that $\dms = m^4/M^2 \, \tilde{s}$, where 
$\tilde{s}$ is a function of the hierarchy parameters 
$\epsilon_D$, $\eta_{1,2}$ and the mass matrix parameters 
$a$, $b$, $g$, $\alpha$ and $\beta$. Its value is 
for $m \simeq v$ and $M \simeq 10^{16}$ GeV located around 10.\\

\subsection{\label{sec:CP}$CP$ Violation in Neutrino 
Oscillation experiments and Neutrinoless Double Beta Decay}
We shall investigate now the predictions of the scenario 
for the $CP$ asymmetries in neutrino oscillation experiments and 
for neutrinoless 
double beta decay and its connection to leptogenesis. 
The interplay between these  
low and high energy parameters has recently been analyzed in a 
number of publications \cite{PPR3,others,FGY1,FGY2,JCP}. Instead of 
trying to identify the low energy Dirac and Majorana phases and express  
them in terms of the available high energy phases in Eq.\ (\ref{eq:md}),  
we shall work as convention--independent as possible. 

\noindent We can calculate the rephasing invariant $CP$ observable 
$J_{CP}$, which can be written as \cite{JCP} 
\be
J_{CP} = -\frac{{\rm Im} (h_{12} \, h_{23} \, h_{31} )}
{\Delta m^2_{21} \, \Delta m^2_{31} \, \Delta m^2_{32}}
~,  \mbox{ where } 
h = m_\nu m_\nu^\dagger ~. 
\ee
With the help of $m_\nu$ given in Eq.\ (\ref{eq:mnu33}) we find with 
the choice of $\epsilon_D^2 \, \eta_1 \simeq 1$ and 
$\eta_1 \simeq 10 \, \eta_2$ that the leading term is given by 
\be
-{\rm Im} (h_{12} \, h_{23} \, h_{31} ) \simeq 
\frac{m^{12}}{M^6} \epsilon_D^4 \, \eta_1 \, \eta_2^4 \, a^2 \, b^2 \, g^2 \, 
(1 + g^2) \, \sin 2 \beta  \simeq 
\frac{2 \, m^{12}}{M^6} \epsilon_D^4 \, \eta_1 \, \eta_2^4 \, a^2 \, b^2
\, \sin 2\beta~.
\ee
With the help of $\Delta m^2_{31} \simeq \Delta m^2_{32} 
\simeq m_3^2 \simeq (2\eta_2 \, m^2/M)^2$ we find with our definition 
for \dms{} that in leading order 
\be \label{eq:JCP}
J_{CP} \simeq \frac{1}{8} \, \epsilon_D^4 \, \eta_1 \, a^2 \, b^2 \, 
\tilde{s} \, \sin 2\beta ~.
\ee
For our representative values we find that 
$J_{CP} \sim 10^{-2} \, a^2 \, b^2 \, \sin 2 \beta$. 
Recall that for, e.g., $\tan^2 \theta_{12} = 0.45$, 
$\sin^2 2 \theta_{23} = 1$ and $\sin^2 \theta_{13} = 0.01$ the 
invariant $J_{CP}$ is given by 
\be
J_{CP} = {\rm Im} \left\{ U_{e1} \, U_{\mu 1}^\ast \, 
U_{e2}^\ast \, U_{\mu 2} \right\} = 
\frac{1}{8} \, \sin 2 \theta_{12}  \, \sin 2 \theta_{23}  
 \, \sin 2 \theta_{13} \, \cos \theta_{13} \, \sin \delta 
\simeq  0.02 \, \sin \delta ~.
\ee
Thus, it is confirmed that $\theta_{13}$ is sizable in the framework under 
study. Since $\dms = |m_2|^2 - |m_1|^2$ depends 
on $\eta_1$, $\eta_2$, $\epsilon_D$, $a$, $b$, $g$, $\alpha$ and $\beta$, 
whereas 
the decay asymmetry is proportional to $\sin 2 \beta$, there is no simple 
connection between the size of $J_{CP}$ and $Y_B$. 
It is seen, however, that --- due to the same dependence on $\beta$ --- 
vanishing $J_{CP}$ is incompatible with successful leptogenesis 
and that $J_{CP}$ has the same sign as the baryon asymmetry. 
The case $\epsilon_D = 0$, i.e., the presence of 
only one Dirac mass, corresponds to 
an effective 2 flavor system in which $J_{CP}$ has to 
vanish, as confirmed by Eq.\ (\ref{eq:JCP}). 

\noindent Finally we 
can analyze the prediction of the scenario for neutrinoless double beta 
decay. From Eq.\ (\ref{eq:mnu33}) and our usual assumptions of the 
parameters 
we find that the absolute values of the $ee$ element of $m_\nu$ is 
\be
\meff \equiv \left|(m_\nu)_{ee}\right| \simeq \frac{m^2}{M} \, 
a^2 \, \epsilon_D^2 \, \eta_2 \simeq 3 \, a \cdot 10^{-4} \, {\rm eV}~.
\ee
Neutrinoless double beta decay triggered by values of \meff{} smaller 
than $10^{-3} \, {\rm eV} $ will probably be unobservable \cite{0vbb}. 
With Eqs.\ (\ref{eq:m3s}) and (\ref{eq:ue3}) 
we can however write an interesting 
correlation of parameters, namely: 
\be
\meff \simeq \sqrt{\dma} \, |U_{e3}|^2~.
\ee

\noindent In summary, the same phase governs the 
$CP$ asymmetry in neutrino oscillations and the decay asymmetry, 
whereas there is no correlation of the leptogenesis phase 
with the effective mass in neutrinoless 
double beta decay. The very same features have been found for the 
minimal see--saw model \cite{FGY1}, which is defined as 
having only 2 heavy Majorana neutrinos and 2 zeros in the Dirac mass matrix. 
Given the presence of two zeros (or very small entries) 
in our $m_D$ (see Eq.\ (\ref{eq:md})) and the fact 
that $M_3 \gg M_{2,1}$, it is very interesting that we encounter the same 
situation. Note however that different variations of the model, 
which have been discussed lately in the literature \cite{FGY2}, 
do not necessarily display the mentioned correlations of the phases.

\subsection{\label{sec:small}Effects of 
entries of order $\epsilon_D^4$ in $m_D$}
The question arises if it is valid to 
neglect terms of order $\epsilon_D^4$ in the 11, 13 and 31 entries 
of $m_D$ in Eq.\ (\ref{eq:md}). We therefore repeat the calculation with 
terms of this order. One finds that new contributions to 
$m_\nu$ are suppressed by one or two orders of $\epsilon_D$. 
Regarding the LFV violating decays, one observes that the term 
$|(m_D m_D^\dagger)_{31}|^2$ now has the leading contribution proportional 
to $\epsilon_D^4 \, h_1$, where $h_1$ is the absolute value of the 13 element 
of $m_D$. The other terms acquire subleading new 
contributions stemming from the new entries in $m_D$.  
Thus, Eq.\ (\ref{eq:BRs}) is modified to 
\be \label{eq:BRsnew}
BR(\mu \ra e \, \gamma) 
\simeq \frac{a^2}{h_1^2} \, \epsilon_D^2 \, BR(\tau \ra e \, \gamma) 
\simeq \frac{a^2}{f^2} \, \epsilon_D^{6} \, 
BR(\tau \ra \mu \, \gamma)~,
\ee
or, numerically:
\be \label{eq:BRsnum}
BR(\mu \ra e \, \gamma) 
\sim 10^{-2} \, BR(\tau \ra e \, \gamma) 
\sim 10^{-6} \, BR(\tau \ra \mu \, \gamma)~.
\ee
Note the analogy of these ratios with the ones presented in 
\cite{PPR3}, where also a hierarchical $m_D$ was assumed. 
One sees that the small entries of order $\epsilon_D^4$ change the 
ratio between $BR(\mu \ra e \, \gamma)$ and 
$BR(\tau \ra e \, \gamma)$ by a factor of $\epsilon_D^2 \simeq 10^{-2}$. 

\noindent The decay asymmetry $\varepsilon_1$ is also 
slightly altered. It reads now 
\be \label{eq:varepsnew}
\varepsilon_1 = \frac{\D 3 \, m^2}{\D 8 \, \pi \, v^2} \, \epsilon_D^2 \, 
\left( \epsilon_D^2 \, \frac{\D \epsilon_{M1}}{\D \epsilon_{M2}} 
\sin 2 \beta  + 
\frac{h_2^2}{b^2} \, \epsilon_{M1} \, \sin 2 \delta_2 \right)~,
\ee
where $h_2$ and $\delta_2$ are the absolute value and phase of the 
31 entry of $m_D$. For $\epsilon_{M1} \ll \epsilon_D^2$, the situation 
we are interested in, we recover the form given in 
Eq.\ (\ref{eq:vareps3}). 
Thus, small entries in $m_D$, which were neglected 
in Eq.\ (\ref{eq:md}), have in our framework some influence on the ratios 
of the LFV violating decay branching ratios but only little influence on 
$m_\nu$ and $\varepsilon_1$.

\section{\label{sec:concl}Conclusions}
The see--saw mechanism with hierarchical Dirac and Majorana neutrino 
masses was reanalyzed in the presence of $CP$ phases. 
A consistent and appealing framework of neutrino mixing 
phenomenology and leptogenesis was found, in which 
one of the heavy Majorana neutrinos is much heavier 
than the other two, which in turn display a mild hierarchy. 
It was investigated how large neutrino mixing can 
be generated starting from hierarchical mass matrices in the see--saw 
mechanism. 
Ratios for the branching ratios of LFV charged lepton 
decays are predicted, which are sensitive to small entries in $m_D$. 
A natural hierarchy of the mixing angles in accordance with 
observation is found and it holds 
$|U_{e3}|^2 \gs 10^{-3}$, which is observable in the not so far future. 
There can be an interplay between too large solar neutrino mixing and 
a too small baryon asymmetry.  
The $CP$ asymmetry in neutrino oscillations 
has the same sign as the 
baryon asymmetry of the universe and successful leptogenesis 
implies non--zero and measurable $J_{CP}$. Neutrinoless 
double beta is not linked with the leptogenesis phase and 
will probably not be observable. The framework under study 
resembles in this respect very much the minimal see--saw model.

\vspace{0.5cm}
\begin{center}
{\bf Acknowledgments}
\end{center}
I thank S.\ Pascoli and S.T.\ Petcov for helpful comments and discussions. 
The hospitality of the Max--Planck Institut 
f\"ur Physik, M\"unchen, where part of this study was performed, is 
gratefully acknowledged. 
This work was supported in part by 
the EC network HPRN-CT-2000-00152.


\newpage

\begin{samepage}
\thispagestyle{empty}
\begin{figure}
\begin{center}
\vspace{-1cm}
\epsfig{file=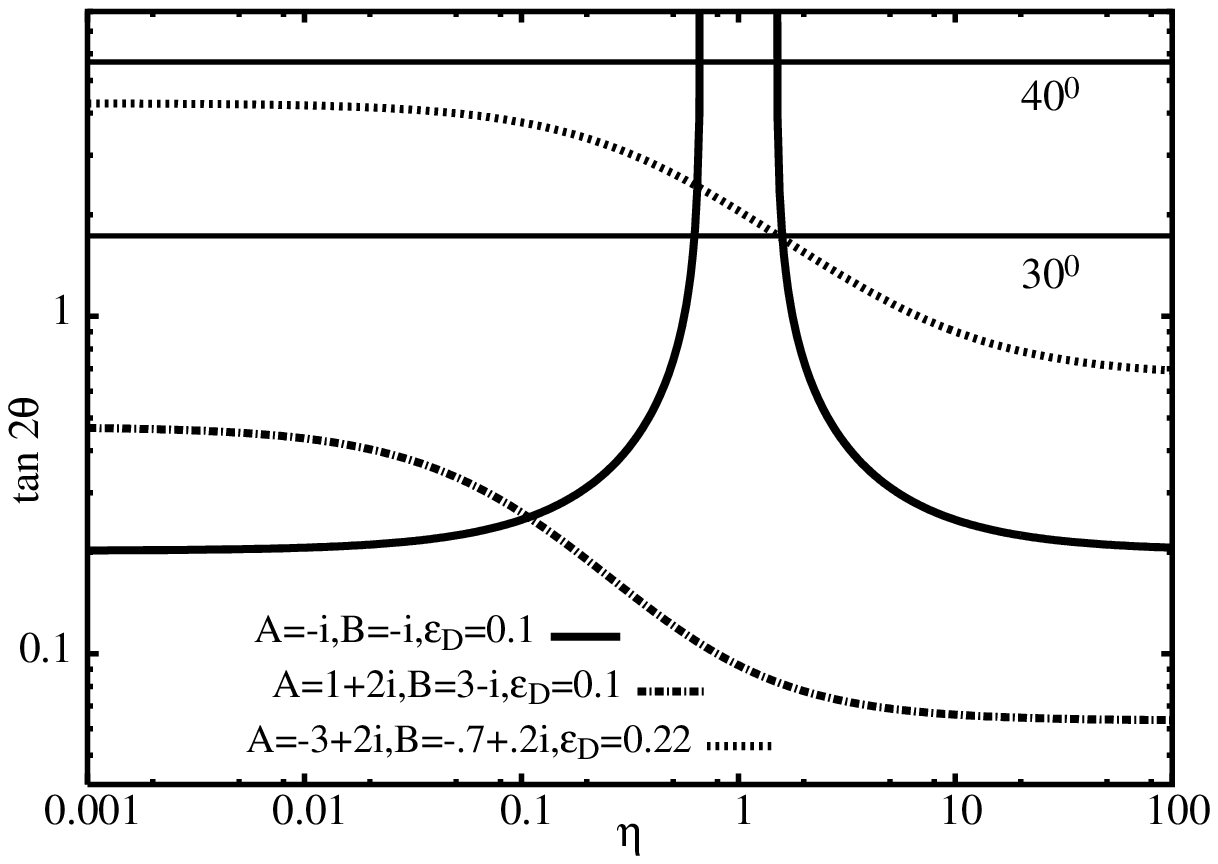,width=13cm,height=8cm}
\caption{\label{fig:examix22}
Result for the mixing angle in a $2 \times 2$ framework, 
Eq.\ (\ref{eq:t2t22}), obtained for 
hierarchical Dirac and Majorana neutrino mass matrices 
$m_D$ and $M_R$ and different values of the relevant parameters. }
\end{center}
\end{figure}

\begin{figure}
\begin{center}
\vspace{-1cm}
\epsfig{file=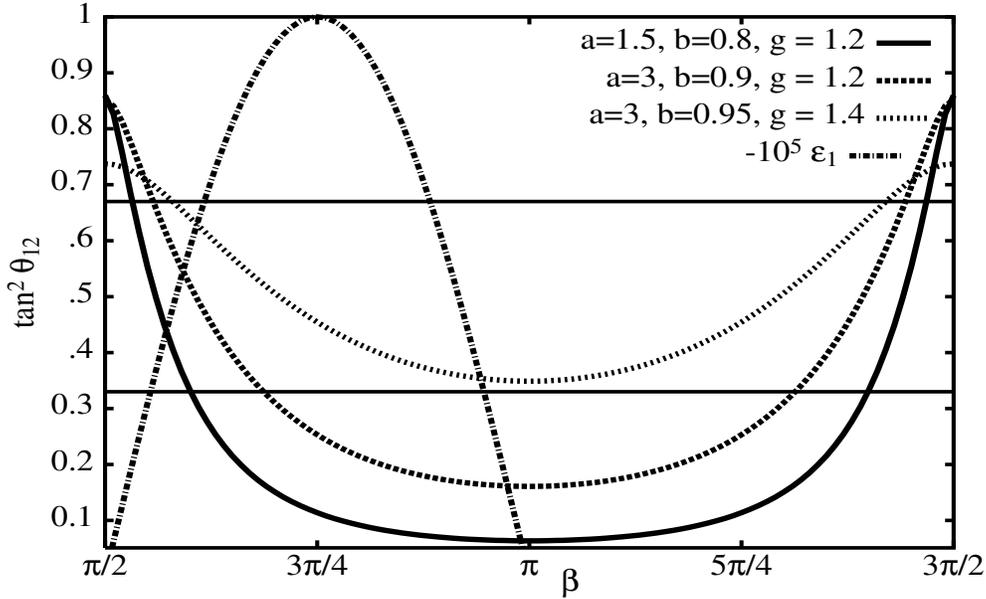,width=13cm,height=8cm}
\caption{\label{fig:examix33}
Result for the mixing parameter 
$\tan^2 \theta_{12}$, 
as obtained from Eq.\ (\ref{eq:t12}),  
for different $a$, $b$ and $g$ 
as a function of the leptogenesis phase $\beta$. 
The range as implied by experiment is indicated. 
The values of $|U_{e3}|^2 $ are 0.009, 0.033 and 0.027, respectively. 
For $g = 1.2$ (1.4) atmospheric neutrino mixing is given by 
$\sin^2 2 \theta_{23} \simeq 0.97$ (0.90).
Plotted is also the 
decay asymmetry $\varepsilon_1$ from Eq.\ (\ref{eq:vareps3}) multiplied with 
$-10^5$ (dash--dotted).}
\end{center}
\end{figure}

\end{samepage}

\clearpage

\setcounter{section}{0}
\renewcommand{\thesection}{\Alph{section}}
\def\sectionname{Appendix}

\section{\label{sec:app33}Diagonalization of a complex and hierarchical 
symmetric $3 \times 3$ matrix}
We present for completeness our formulae for the 
diagonalization of a complex and hierarchical symmetric 
$3 \times 3$ matrix. It is a special case of 
the general strategy as outlined, e.g., in Ref.\ \cite{king}. 
In the diagonalization of a $2 \times 2$ matrix three phases were present. 
We saw that two of them can be absorbed in the charged lepton fields. 
Diagonalizing a complex $3 \times 3$ matrix through three 
consecutive $2 \times 2$ diagonalizations will introduce 6 phases, 
which in principle can influence the mixing angles. 
In our case, however, they do not. 
We take advantage of the somewhat more simple structure of $m_\nu$ in the 
hierarchical situation we consider. 
It is convenient to express the results in terms of mixing angles. 
Regarding the phases, as stated in the text, 
we prefer not to identify the low energy Dirac and Majorana phases but 
work with convention independent quantities like $J_{CP}$.  
Consider a symmetrical neutrino mass matrix  
\be 
m = 
\left( 
\bad 
a & b & d \\[0.3cm]
\cdot & e & f \\[0.3cm]
\cdot & \cdot & g 
\ea
\right)~,
\ee
where the 23 block has entries larger than the other elements. 
The strategy outlined in \cite{king} is to first rephase 
the mass matrix with 
$P_2 \, m \, P_2$, where $P_2$ is a diagonal phase matrix with 
complex entries on the 22 and 33 elements. 
Then, one puts zeros in the 23 and 13 elements of $m$ by diagonalizing first 
the 23 submatrix and then the resulting 13 submatrix. 
Then the matrix is again rephased by a diagonal 
phase matrix containing only one complex entry on the $22$ element. 
After that, we have to diagonalize the 12 submatrix and end up in 
this way with a diagonal matrix. The eigenstates are however still 
complex. Thus, by again rephasing the diagonal matrix and absorbing 
these three phases in the charged leptons, we end up with the desired 
three real diagonal entries, three mixing angels and three phases. 

\noindent In our case, the 23 submatrix of Eq.\ (\ref{eq:mnu33}) 
is effectively real, since we choose $\eta_2 \simeq 10$. 
Therefore, the first rephasing with $P_2$ is not necessary and 
there is also no phase in the 23 rotation. Thus, 
the 23 submatrix is diagonalized via 
$R_{23}^T \, m \, R_{23}$ where 
\be
R_{23} = 
\left( 
\bad 
1 & 0 & 0 \\[0.3cm]
0 & c_{23} & s_{23} \\[0.3cm]
0 & - s_{23} & c_{23}
\ea
\right)~,
\ee
where $c_{23} = \cos \theta_{23}$ and 
$s_{23} = \sin \theta_{23}$. 
The resulting matrix $m'$ is 
\be
m' = 
\left( 
\bad 
a & b \, c_{23} - d \, s_{23} & b \, s_{23} + d \, c_{23} \\[0.3cm]
\cdot & m_2' & 0  \\[0.3cm]
\cdot & \cdot & m_3' 
\ea
\right) \equiv 
\left( 
\bad 
a & b' & d' \\[0.3cm]
\cdot &  m_2' & 0 \\[0.3cm]
\cdot & \cdot & m_3'  
\ea
\right)~,
\ee
for 
\be
m_{2,3}' = \frac{1}{2} \, 
\left( (e + g) \mp \sqrt{(e - g)^2 + 4 \, f^2} \right)
\ee
and 
\be
\tan 2 \theta_{23} = \frac{2 \, f}
{g\, - e } ~.
\ee
Now the 13 submatrix of $m'$ is diagonalized via 
$R_{13}^T \, m' \, R_{13}$ with 
\be
R_{13} = 
\left( 
\bad 
c_{13} & 0 & s_{13} \\[0.3cm]
0 & 1  & 0  \\[0.3cm]
- s_{13}^\ast & 0 & c_{13}
\ea
\right)~,
\ee
where $c_{13} = \cos \theta_{13}$ and 
$s_{13} = \sin \theta_{13} \, e^{i \phi_{13}}$. 
The resulting matrix $m''$ reads 
\be
m'' = 
\left( 
\bad 
m_1'' & b' \, c_{13}  & 0  \\[0.3cm]
\cdot & m_2' & b' \, s_{13}  \\[0.3cm]
\cdot & \cdot & m_3'' 
\ea
\right) \simeq 
\left( 
\bad 
m_1'' & b' & 0  \\[0.3cm]
\cdot & m_2' & 0 \\[0.3cm]
\cdot & \cdot & m_3'' 
\ea
\right)~,
\ee
where the last approximation takes into account the smallness of 
$\theta_{13}$ as implied by the reactor experiments 
and the hierarchical structure of $m$. 
The masses and the mixing angle are given by 
\be
m_{1,3}'' = \frac{1}{2} \, 
\left( (a + m_3') \mp \sqrt{(a - m_3')^2 + 4 \, d'^2} \right)
\ee
and 
\bea
\tan 2 \theta_{13} = \frac{\D 2 \, d'}{\D m_3' \, e^{-i \phi_{13}} - 
a \, e^{i \phi_{13}}} 
\simeq \frac{\D 2 \, d' \, e^{i \phi_{13}}}{\D m_3'}~,\\[0.5cm]
\mbox{ where } \, 
\arg (d') =   
\arg(m_3' \, e^{-i \phi_{13}} - a \, e^{i \phi_{13}}) \Rightarrow 
\phi_{13} \simeq \arg(m_3') - \arg(d')~.
\eea
From Eq.\ (\ref{eq:mnu33}) we see that the 11 element 
of our $m_\nu$ (here called $a$) 
is much smaller than $m_3'$ as given in Eq.\ (\ref{eq:m3s}). 
The phase $\phi_{13}$ is therefore 
suppressed and does not influence the magnitude of $\theta_{13}$. 
The eigenvalue $m_3'' \equiv m_3$ is already the heaviest 
eigenvalue of the matrix $m$. 
Now we rephase $m''$ through a diagonal phase matrix $P$ with only the 
22 entry being complex, $P = {\rm diag}(1,e^{i \phi},1)$. 
Finally, the 12 submatrix of $m''$ gets diagonalized by 
$R_{12}^T \, m'' \, R_{12}$ where 
\be
R_{12} = 
\left( 
\bad 
c_{12} & s_{12} & 0 \\[0.3cm]
- s_{12}^\ast & c_{12}  & 0\\[0.3cm]
0 &  0 & 1
\ea
\right) 
\ee
and for the masses and mixing angle holds  
\be
m_{1,2} = \frac{1}{2} \, 
\left( (m_1'' + m_2') \mp \sqrt{(m_1'' - m_2')^2 + 4 \, b'^2} \right)
\ee
as well as 
\be
\tan 2 \theta_{12} = \frac{2 \, b' \, e^{i \phi}}
{m_2'  \, e^{-i \phi_{12}} \, e^{2 i \phi} - 
m_1''  \, e^{i \phi_{12}}} \, 
\mbox{, where } \, \arg (b'\, e^{i \phi}) = 
\arg(m_2' \, e^{-i \phi_{12}} \, e^{2 i \phi} - 
m_1''  \, e^{i \phi_{12}})~.
\ee
In our case it turns out that $m_2' \gg m_1''$, therefore $\phi$ and 
$\phi_{12}$ do not influence the magnitude of $\theta_{12}$. 
The mass states are in general still complex. 
Rephasing these states through a diagonal phase matrix and absorbing them in 
the charged lepton fields 
then leaves us with the correct number of 
three phases in $U$.

\end{document}